\documentstyle[12pt,epsf]{article}
\makeatletter

\@addtoreset{equation}{section}
\makeatother

\newcommand{\bm}[1]{\mbox{\boldmath$#1$}}
\newcommand{\non}{\nonumber}
\def\ie{{\it i}.{\it e}.}
\def\eg{{\it e}.{\it g}.}
\def\L{{\rm L}}
\def\R{{\rm R}}

\def\GSC{G_{{\rm SC}}}
\def\PS{{\rm PS}}
\def\F{{\rm F}}
\def\fun#1{\!\left(#1\right)}
\def\wb#1{\vbox{\ialign{##\crcr%
          \hskip 1.0pt\hrulefill\hskip 0.3pt%
          \crcr\noalign{\kern-1pt\vskip0.07cm\nointerlineskip}%
          $\hfil\displaystyle{#1}\hfil$\crcr}}}
\def\abs#1{\left| #1 \right|}

\def\wt#1{{\widetilde #1}}
\def\VEV#1{\left<#1\right>}
\def\order#1{{\cal O}\!\left(#1\right)}
\newcommand{\gsim}{\ %
{\raisebox{0.4ex}{${\textstyle >}$}}{\kern-0.77em}
{\raisebox{-0.7ex}{${\textstyle \sim}$}}\ }
\newcommand{\lsim}{\ 
{\raisebox{0.4ex}{${\textstyle <}$}}{\kern-0.75em}
{\raisebox{-0.7ex}{${\textstyle \sim}$}}\ }
\def\alp#1{\alpha_{#1}^{\phantom{2}}}
\def\flipped{1}		%\def\flipped{\tilde{1}}
\def\alphaFlipped#1#2{\alpha_{#1}^{\prime{#2}}}
\def\alphafive#1{\alphaFlipped{5}{#1}}
\def\alphaone#1{\alphaFlipped{1}{#1}}
\def\MFlipped#1#2{M_{#1}^{\prime{#2}}}
\def\Mfive#1{\MFlipped{5}{#1}}
\def\Mone#1{\MFlipped{1}{#1}}
\def\ratio{r}
\def\bFlipped#1{b'_{#1}}
\def\bfive{\bFlipped{5}}
\def\bone{\bFlipped{1}}
\def\CFlipped#1{C'_{#1}}
\def\Cfive#1{\CFlipped{{#1}5}}
\def\Cone#1{\CFlipped{{#1}1}}
\begin{document}
\begin{titlepage}
\vspace*{-20mm}
 \begin{flushright}
 \parbox{0.22\textwidth}{
  KANAZAWA-02-01\\[-1mm]
  KUNS-1751\\[-1mm]
  NIIG-DP-02-01}
 \end{flushright}
\begin{center}
   
    {\Large\bf 
    Sfermion Mass Degeneracy, Superconformal Dynamics\\[5pt]
    and Supersymmetric Grand Unified Theories
    } 

\vspace*{1cm}
{%\large
    Tatsuo~{\sc Kobayashi}\rlap,\,\footnote{
	E-mail: kobayash@gauge.scphys.kyoto-u.ac.jp}
    Hiroaki~{\sc Nakano}\rlap,\,\footnote{
	E-mail: nakano@muse.hep.sc.niigata-u.ac.jp}
    Tatsuya~{\sc Noguchi}\rlap,\,\footnote{
	E-mail: noguchi@gauge.scphys.kyoto-u.ac.jp}
and
    Haruhiko~{\sc Terao}\,\footnote{
	E-mail: terao@hep.s.kanazawa-u.ac.jp}
}

\vspace*{5mm}
  $^{*\ddagger}${\it 
  Department of Physics, Kyoto University, Kyoto 606-8502, Japan}\\
  $^{\dagger}${\it 
  Department of Physics, Niigata University, Niigata 950-2181, Japan}\\
  $^{*\ddagger}${\it 
  Institute for Theoretical Physics, Kanazawa University, 
  Kanazawa 920-1192, Japan}\\
\vspace{0.8cm}
\end{center}
  
\begin{abstract}
We discuss issues in a scenario that
hierarchical Yukawa couplings are generated 
through strong dynamics of superconformal field theories (SCFTs).
Independently of mediation mechanism of supersymmetry breaking,
infrared convergence property of SCFTs can provide 
an interesting solution to supersymmetric flavor problem;
sfermion masses are suppressed around the decoupling scale of SCFTs
and eventually become degenerate to some degree, 
thanks to family-independent radiative corrections 
governed by the SM gaugino masses.
We discuss under what conditions the degeneracy of sfermion mass 
can be estimated in a simple manner. 
We also discuss the constraints from lepton flavor violations.
We then study explicitly sfermion mass degeneracy 
within the framework of grand unified theories coupled to SCFTs. 
It is found that the degeneracy for right-handed sleptons
becomes worse in the conventional $SU(5)$ model than in the MSSM.
On the other hand, in the flipped $SU(5) \times U(1)$ model, 
each right-handed lepton is still an $SU(5)$-singlet,
whereas the bino mass $M_1$ is determined
by two independent gaugino masses of $SU(5) \times U(1)$.
These two properties enable us to have 
an improved degeneracy for the right-handed sleptons.
We also speculate how further improvement can be obtained 
in the SCFT approach.
\end{abstract}
\end{titlepage}

\setcounter{footnote}{0}
\setcounter{equation}{0}
%%%%%%%%%%%%%%%%%%%%%%%%%%%%%%%%%%%%%%%%%%%%%%%%%%%%%%%%%%%%%%%%%%%%%%
\section{Introduction}
\label{sec:intro}
%%%%%%%%%%%%%%%%%%%%%%%%%%%%%%%%%%%%%%%%%%%%%%%%%%%%%%%%%%%%%%%%%%%%%%

Understanding the origin of hierarchical fermion masses and 
mixing angles is one of most important issues in particle physics.
The Froggatt-Nielsen mechanism is a famous mechanism 
to realize hierarchical Yukawa couplings~\cite{FN,IR}. Recently 
new ideas related to extra dimensions have also been discussed.

In models with softly-broken supersymmetry (SUSY), 
a mechanism that generates the hierarchical structure of 
Yukawa couplings generally affects the sfermion sector; 
%\ie, such mechanism would lead to a characteristic pattern of 
one would have a characteristic pattern of 
sfermion masses and SUSY-breaking trilinear couplings.
{}For example, the Froggatt-Nielsen mechanism with an extra $U(1)$ 
gauge symmetry leads to the so-called $D$-term contribution 
to soft scalar masses\rlap,\,\footnote{
See Ref.~\cite{nonanomalous} 
for $D$-term contributions through GUT breaking 
and Ref.~\cite{anomalous} for anomalous $U(1)$ breaking.
}\,which are proportional to the charges under the broken $U(1)$ symmetry.
%that is, the $D$-term contributions are flavor-dependent.
Such a pattern could be tested
if superpartners will be discovered and sfermion masses as well as 
trilinear scalar couplings will be measured in future experiments.
Even at present, 
soft SUSY-breaking parameters are constrained rather severely 
from the exprimental bounds on flavor changing neutral current 
(FCNC) processes as well as CP violation.
This is the SUSY flavor problem. 
For instance, the flavor-dependent $D$-term contributions are 
generically dangerous. 
In general, the SUSY flavor problem can be solved 
if either of the following three is realized
at least for the first two families;
1)~diagonal and degenerate sfermion masses,
2)~decoupling of heavy sfermions, and 
3)~the alignment between the fermion and sfermion bases.
Much effort has been devoted to realize the first solution
by seeking a flavor-blind mediation mechanism of SUSY breaking.

Nelson and Strassler~\cite{NS1} have recently proposed 
an interesting mechanism to realize hierarchical Yukawa couplings.
The setup is the minimal supersymmetric standard model (MSSM),
or its extention, coupled to superconformal (SC) sector.
The SC sector is strongly coupled and assumed to have 
an infrared (IR) fixed point~\cite{BaZa,seiberg}.
The first and second families of quarks and leptons gain 
a large and positive anomalous dimensions through the SC dynamics.
Then their Yukawa couplings to electroweak Higgs fields are
suppressed hierarchically at the scale $M_C$ where 
the SC sector is assumed to decouple from the MSSM sector.

The SC fixed point has more interesting consequences.
When pure superconformal field theory (SCFT) 
is perturbed by soft SUSY-breaking terms,
general argument shows that
such perturbation is exponentially suppressed 
toward the SC fixed point~\cite{kkkz,lr,KT,NS2}.
Specifically one expects that a sfermion mass is suppressed
at the decoupling scale $M_C$ 
and eventually receives radiative corrections 
governed by the SM gaugino masses, which are flavor-blind.
Hence, we would have degenerate mass spectrum of sfermions
like the `no-scale' model (at least for the first two families).
This possibility has already been mentioned in Ref.~\cite{NS1}.
In this scenario, soft scalar masses are to be controlled by 
the flavor mechanism that generates hierarchical Yukawa couplings, 
no matter how SUSY breaking is mediated and no matter what 
initial conditions of soft SUSY-breaking terms are.
This approach, which we shall pursue in this paper,
is quite opposite to the usual scenario in which
degenerate soft scalar masses are supposed to be derived 
by a flavor-blind mediation of SUSY breaking.

When the pure SCFT is perturbed by
the SM gauge interactions and the SM gaugino masses,
each sfermion mass is not completely suppressed,
but converges on a flavor-dependent value~\cite{KT,NS2}.
The convergent value is one-loop suppressed
since the SM gauge couplings are perturbatively small.
It is then plausible that
sfermion masses at the weak scale may be calculated solely 
in terms of the SM gauge coupling constants and gaugino masses.
In fact, under some assumptions,
we can estimate the degeneracy factor 
$\Delta_{\tilde{f}}$ of sfermion masses 
up to a single model-dependent parameter $\Gamma_i$~\cite{KT}.
It was found that for $M_C > 10^{10}$~GeV, the degeneracy factor 
is $0.005 - 0.01$ for squarks and $0.05 - 0.1$ for sleptons.
In particular, 
the right-handed sleptons are not well degenerate in mass.
Note that these degeneracy factors are evaluated 
in the sfermion basis.

In this paper,
we first examine the assumptions that are implicit
in the estimation of sfermion mass degeneracy
in the present SCFT approach.
We also discuss to what extent the degeneracy is required 
in the Nelson-Strassler scenario.
To this aim, we take into account the fact that
FCNC processes, when correctly evaluated in the fermion basis, 
have additional suppression 
since the Nelson-Strassler mechanism 
leads to hierarchical Yukawa matrices.

It is natural to extend the analysis to grand unified theories (GUTs) 
and to examine how much degeneracy of sfermion masses is achieved 
by coupling to SCFTs. 
That is our second purpose; 
we explicitly study sfermion mass degeneracy 
within the framework of GUTs coupled to SC sectors.
We take the $SU(5)$ and the flipped $SU(5)\times U(1)$ 
as a prototype of GUT models.
We shall show how much degeneracy of sfermion masses 
is expected in each case. It turns out that
a simple extention of the MSSM to the $SU(5)$ makes
the degeneracy of the right-handed sleptons worse,
because each right-handed lepton is embedded 
into a higher-dimensional representation.
The situation is different in the flipped $SU(5)\times U(1)$ case,
since the right-handed leptons remain $SU(5)$-singlets~\cite{Flip1}, 
and the bino mass $M_1$ is determined by a combination of 
two independent gaugino masses of $SU(5)\times{}U(1)$.

This paper is organized as follows.
After a brief review on the Nelson-Strassler mechanism 
in section~\ref{subsec:Yukawa},
we outline in subsection~\ref{subsec:mass} 
the results of Ref.~\cite{KT} on 
degeneracy of sfermion masses in the sfermion basis 
within the framework of the MSSM coupled to SC sectors.
We also discuss under what conditions
the degeneracy factor can be estimated in a simple mannar.
The FCNC constraints are examined in subsection~\ref{subsec:FCNC}
by taking into account Yukawa-diagonalizing matrices.
In sections~\ref{sec:GG} and \ref{sec:flipped}, 
we extend the model to GUTs and investigate generic features.
Specifically we argue that
the degeneracy of the right-handed sleptons becomes worse 
in $SU(5)$ models than the MSSM.
We then show how the degeneracy for the right-handed sleptons 
can be improved in the flipped $SU(5)\times U(1)$ models.
In section~\ref{sec:threshold}, 
we briefly discuss various sorts of threshold effects,
which might affect the previous results.
This in particular includes
$D$-term contributions on sfermion masses 
in the flipped $SU(5)\times U(1)$ models.
Section~\ref{sec:conclusion} is devoted to conclusion and discussion.

We note that Luty and Sundrum~\cite{LutySundrum} discuss 
a scenario of `conformal sequestering\rlap,'\,
which is another interesting approach 
to the SUSY flavor problem based on four-dimensional SCFTs.
Although closely related to the present SCFT approach,
it is slightly different in that the SCFT couples 
to messenger fields of SUSY breaking, not to quarks and leptons.

%%%%%%%%%%%%%%%%%%%%%%%%%%%%%%%%%%%%%%%%%%%%%%%%%%%%%%%%%%%%%%%%%%%%%%
\section{MSSM coupled to SC sector}
\label{sec:MSSM}
%%%%%%%%%%%%%%%%%%%%%%%%%%%%%%%%%%%%%%%%%%%%%%%%%%%%%%%%%%%%%%%%%%%%%%

\subsection{Nelson-Strassler mechanism for Yukawa hierarchy}
\label{subsec:Yukawa}

Here we give a brief review on 
the Nelson-Strassler (NS) mechanism that generates 
the hierarchical structure of Yukawa couplings~\cite{NS1}.
We assume two sectors: 
One is the SM sector and the other is the SC sector. 
The SM sector has the gauge group $SU(3)\times SU(2) \times U(1)_Y$,
or its extention, and contains three families of quarks and leptons 
$\psi_i$ $(i=1,2,3)$ as well as Higgs fields $H$.
The SC sector has a gauge group $\GSC$ and matter fields $\Phi_r$.
The fields $\psi_i$ and $H$ are taken to be $\GSC$-singlets.
The following superpotential is assumed;
\begin{equation}
W = y_{ij\,}\psi_i\psi_j{}H
   +\lambda_{irs\,}\psi_{i\,}\Phi_r\Phi_s+\cdots \ ,
\label{messenger}
\end{equation}
where the first term describes the ordinary Yukawa couplings 
in the SM sector and the ellipsis denotes 
terms including only $\Phi_r$.
The second term represents the couplings 
of quarks and leptons $\psi_i$ to the SC sector, 
which we refer to as {\it messenger couplings}\/.
{}For the messenger coupling $\lambda_{irs}$ to be allowd 
by gauge invariance, either of $\Phi_r$ or $\Phi_s$ should 
belong to a nontrivial representation under the SM gauge group.

With sufficiently many matter fields, 
the $\GSC$ gauge theory resides in `conformal window';
the SC gauge coupling $g'$ has an IR fixed point~\cite{BaZa,seiberg}.
Suppose that the messenger couplings $\lambda_{irs}$ 
as well as $g'$ approach IR fixed points.
At this new fixed point, the field $\psi_i$ gains
a large anomalous dimension $\gamma_i^* = \order{0.1\,\hbox{--}\,1}$ 
through superconformal dynamics. As a result, the Yukawa couplings 
in the SM sector obey the power law and behave roughly like
\begin{equation}
y_{ij}(M_C)\,\approx\,
y_{ij}(M_0)\left({M_C \over M_0}\right)^{\gamma_i^*+\gamma_j^*} \ .
\label{yukawa}
\end{equation}
Here $M_0$ is the cut-off scale, 
at which $y_{ij}(M_0) =\order{1}$ is expected. 
Thus the hierarchical structure of Yukawa couplings can be generated
by family-dependent anomalous dimensions $\gamma_i^*$.
The resultant Yukawa matrices are similar to the ones obtained by
the Froggatt-Nielsen (FN) mechanism;
large anomalous dimension in the NS mechanism
corresponds to $U(1)$ charge in the FN mechanism.
As stressed in Ref.~\cite{NS1},
the unitarity of the SCFT guarantees that
the anomalous dimensions $\gamma_i^*$ at the SC fixed point 
are always non-negative, whereas the non-negativity of $U(1)$ charges 
is just the assumption in the conventional FN mechanism.

Since the top Yukawa coupling should not be suppressed, 
the top quark as well as the up-type Higgs field 
must not couple to the SC sector.
Although the bottom quark and tau lepton 
as well as the down-type Higgs could couple, 
we will mainly consider, in what follows, the models in which
only the first two families couple to the SC sector.

\subsection{Degeneracy of sfermion masses}
\label{subsec:mass}

Next we outline, following Ref.~\cite{KT},
how convergent values of soft scalar masses and 
sfermion mass degeneracy can be esitmated.
See also Ref.~\cite{Terao} for a review.
We also discuss some subtleties about such estimation.

The suppression of sfermion masses in the SCFT approach
follows from a general property of 
the renormalization group equations (RGEs)
of soft SUSY-breaking parameters~\cite{softbeta,Terao}.
Let us concentrate for a moment on the diagonal elements 
$m^2_i$ of a sfermion mass-squared matrix
(in a flavor basis in which 
fermion Yukawa matrix takes the form (\ref{yukawa})).
Near an IR attractive fixed point of pure SCFT,
the RGE of $m_i^2$ takes the form
\begin{equation}
\mu\frac{dm^2_{i}}{d\mu}
 = {\cal M}_{ij\,}m^2_{j} \ ,
\label{RGE:pure}
\end{equation}
where the coefficient matrix ${\cal M}_{ij}$ 
encodes full effects of the SC dynamics,
and can be calculated by use of the `Grassmanian expansion method' 
if the anomalous dimension $\gamma_i$ 
is known as a function of coupling constants 
$g'$ and $\lambda_{irs}$. 
This matrix is positive-definite (non-negative)
since the fixed point is IR attractive.
It follows that certain combinations of $m_i^2$
are exponentially suppressed.
Moreover, {\it each}\/ $m^2_i$ is suppressed
if its anomalous dimension $\gamma_i^*$ 
is uniquely determined~\cite{NS1,KT}
by fixed point equations $\beta_{g'}=\beta_{\lambda}=0$.
We assume that in each model considered below, 
there exists a proper set of couplings that satisfies this condition.

When we switch on the gauge couplings 
$\alp{a} = g_a^2/8\pi^2$ and gaugino masses $M_a$ 
in the SM sector ($a=1,2,3$), 
the RGE (\ref{RGE:pure}) gets modified, at the leading order, into
\begin{equation}
\mu\frac{dm^2_{i}}{d\mu}
 = {\cal M}_{ij\,}m^2_{j}
  -\sum_{a=1,2,3} C_{ia\,}\alp{a} M^2_a \ ,
\label{rg-scalar-m}
\end{equation}
where $C_{ia}=4C_2(R_{ia})$ with the quadratic Casimir coefficient 
$C_2(R_{ia})$ of the $R_{ia}$ representation,
and we have neglected for simplicity 
a possible contribution from the $U(1)_Y$ Fayet-Ilipoulos term
${\cal S}\equiv {\rm Tr }\left({Y}m_{i}^2\right)$.
Eq.~(\ref{rg-scalar-m}) implies that
each sfermion mass eventually converges on 
one-loop suppressed value of the order $\alp{a}M_a^2$.
The convergent values generally depend 
on detailed structure of the SC sector,
since the above RGEs are coupled equations for soft scalar masses 
in the SC sector as well as those of squarks and sleptons.

We are interested in 
the convergent value of squark and slepton masses 
at the scale $M_C$ where the SC sector decouples.
Let $m_{\tilde{f}i}$ denote soft scalar mass of
the $i$-th family of squark or slepton $\widetilde{f}$. 
Then the convergent value of $m^2_{\tilde{f}i}$ at $M_C$
can be written in the form~\cite{KT}
\begin{eqnarray}
  m^2_{\tilde f i}\ \longrightarrow\ 
      \frac{1}{\Gamma_{\tilde f i}}
      \sum_{a=1,2,3} C_{fa\,}\alp{a}\fun{M_C}M^2_a\fun{M_C} \ .
\label{conv-scalar-m}
\end{eqnarray}
In this expression,
$C_{fa}$ is the Casimar factor for $f=Q,u,d,L,e$,
and the prefactor $\Gamma^{-1}_{\tilde f i}$ 
summarizes the structure of the SC sector.
Typically, 
we find $\Gamma_i\sim\gamma_i^* =\order{0.1\,\hbox{--}\,1}$
for squarks and sleptons\rlap.\,\footnote{
This is not always true for sfermions in SC sector.
}
If $\gamma_i^*$ and $\gamma_j^*$ are different from each other, 
the factors $\Gamma_{\tilde f i}$ and $\Gamma_{\tilde f j}$ 
are also different.
Only the prefactor $\Gamma^{-1}_{\tilde f i}$ has flavor-dependency.
Consequently, the difference between the first and second families 
is given by
\begin{eqnarray}
  m^2_{\tilde f 2}(M_C) - m^2_{\tilde f 1}(M_C) 
= \left(\frac{1}{\Gamma_{\tilde f 2}} 
       -\frac{1}{\Gamma_{\tilde f 1}}\right)
  \sum_{a=1,2,3} C_{{f}a\,}\alp{a}\fun{M_C}M^2_a\fun{M_C} \ ,
\label{diff-12}
\end{eqnarray}
which is also one-loop suppressed compared with $M_a^2(M_C)$.

We have considered only 
the diagonal elements of sfermion mass-squared matrix.
However, as was shown in Ref.~\cite{NS2} for pure SCFT,
off-diagonal elements are also exponentially suppressed.
Even when we switch on SM effects, 
the RGEs of off-diagonal elements, unlike Eq.~(\ref{rg-scalar-m}),
contain no contribution from the SM gaugino masses 
at the leading order.
Thus the off-diagonal elements converge on sufficiently small values,
which we can safely neglect.

In Ref.~\cite{KT}, sfermion mass degeneracy was examined 
by assuming the MSSM field content below $M_C$.
Here let us recall some results for later convenience.
The sfermion mass receives radiative correction 
$\Delta m_{\tilde f i}^2$ between $M_C$ and $M_Z$, 
which is evaluated to be
\begin{eqnarray}
\label{mssmmassdiff}
\Delta m^2_{\tilde f i}(M_C \rightarrow M_Z)
 &\!=\!& \sum_{a=1,2,3} a_{f a\,}I_a\fun{M_Z,M_C} M^2_a(M_C) \ , 
\label{rad-cor}\\[3pt]
I_a(M_Z,M_C) &\!\equiv\!& 
     1 - \frac{\alpha^2_a(M_Z)}{\alpha^2_a(M_C)} \ , \qquad
a_{fa} \equiv \frac{C_{fa}}{2b_a} \ .
\end{eqnarray}
Here $b_a$ are 
the MSSM gauge beta-function coefficients,
%\rlap,\,\footnote{
%We have adapted the $SU(5)$-motivated normalization
%for the $U(1)$ gauge coupling, \ie,
%$\alp{Y}\equiv\left(3/5\right)\alp{1}$.}
and the factors $a_{fa}$ ($a=1,2,3$) are shown 
in the second, third and fourth columns of 
Table~\ref{table:coefficient} for each matter field.
These radiative corrections are much larger than 
the convergent value $m_{\tilde f i}^2(M_C)$. 
It follows that
\begin{eqnarray}
\left.\frac{m^2_{\tilde f 2} - m^2_{\tilde f 1}}
           {m^2_{\tilde f 2} + m^2_{\tilde f 1}}\right\vert_{M_Z}\!\!
 &=& \frac{1}{2}
     \left(\frac{1}{\Gamma_{\tilde f 2}} 
          -\frac{1}{\Gamma_{\tilde f 1}}
     \right) \Delta_{\tilde  f} \ ,
\end{eqnarray}
where we define the degeneracy factor $\Delta_{\tilde{f}}$ by
\begin{eqnarray}
\Delta_{\tilde  f}
&\!=\!&\frac{\sum_a C_{{\bar f}a\,}\alp{a}\fun{M_C}M^2_a\fun{M_C}}
            {\sum_a a_{f a\,}I_a\fun{M_Z,M_C}M_a^2(M_C)} \ .
\end{eqnarray}
The factor $\Delta_{\tilde{f}}$ serves 
as an estimate of how much degeneracy of sfermion masses 
is achieved in the present framework.
Moreover, it is a calculable quantity
independently of detailed structure of the SC sector,
especially when the SM gaugino masses satisfy 
the `GUT' relation, $M_a/\alpha_a=$ constant for $a=1,2,3$.
For example, taking $M_C = 10^{16}$~GeV gives~\cite{KT}
\begin{eqnarray}
\Delta_{\tilde Q} &\!=\!& 8 \times 10^{-3}\ , \qquad 
\Delta_{\tilde u}  \,=\, 
\Delta_{\tilde d}  \,=\,  6 \times 10^{-3} \ , 
\non\\
\Delta_{\tilde L} &\!=\!& 5 \times 10^{-2}\ , \qquad\qquad\quad
\Delta_{\tilde e\,}\,=\,  1 \times 10^{-1} \ .
\label{Delta:MSSM}
\end{eqnarray}
Unfortunately, the degeneracy factor $\Delta_{\tilde{e}}$ 
for the right-handed slepton is rather large.
The primary reason is that the radiative correction 
to the right-handed sleton mass, which involves only $M_1$, 
is smaller than the others\rlap.\,\footnote{
The degeneracy factor $\Delta_{\tilde{e}}$ 
for the right-handed sleptons can be somewhat reduced~\cite{KT} 
if there is a sizable contribution from 
the $U(1)_Y$ Fayet-Iliopoulos term ${\cal S}$ (with a suitable sign).
Note that we generally expect ${\cal S}\neq 0$
since soft scalar masses of the third family and Higgs fields 
are not constrained by SC dynamics. 
}
However, this result does not necessarily imply that
the present SCFT approach to Yukawa hierarchy leads to
significant lepton flavor-violation,
because the actual size of lepton flavor-violating processes
depends on an explicit form of lepton Yukawa matrix,
as we discuss below.

\begin{table}[tb]
  \begin{center}
    \leavevmode
\begin{tabular}[t]{|c||c|c|c|c|c|}
\hline
\makebox[10mm]{$f$} &
\makebox[18mm]{$a_{f3}$} &
\makebox[18mm]{$a_{f2}$} & 
\makebox[18mm]{$a_{f1}$} & 
\makebox[30mm]{rep.\,under $SU(5)$} &
\makebox[18mm]{$C_{f5}$} \\
\hline\hline
$Q$ & ${{}-8/9\ }$ & ${3/2}$ & ${1/198}$ & $\bm{10}$ & ${72/5}$ \\
\hline
$u$ & ${{}-8/9\ }$ & 0 & ${8/99}$ & $\bm{10}$ &${72/5}$ \\
\hline
$d$ & ${{}-8/9\ }$ & 0 & ${2/99}$ & $\wb{\bm{5}}$ & ${48/5}$ \\
\hline
$L$ & 0 & ${3/2}$ & ${1/22}$ & $\wb{\bm{5}}$ & ${48/5}$ \\
\hline
$e$ & 0 & 0 & ${2/11}$ & $\wb{\bm{10}}$ & ${72/5}$ \\
\hline
\end{tabular}
\caption{Group-theoretical factors for $\Delta_{\bar f}$.
Our normalization convention for the $U(1)$ gauge coupling 
is the $SU(5)$-motivated one,
$\alpha_{1}\equiv\left(5/3\right)\alpha_{Y}$.
}
\label{table:coefficient}
  \end{center}
\end{table}

Some remarks are to be added here.
The convergent value (\ref{conv-scalar-m}) is determined 
by the SM one-loop term in the RGE~(\ref{rg-scalar-m}).
In general, 
{\it higher-loop terms like
$\left(\lambda^2_*/8\pi^2\right)^n\times\alp{a}M_a^2$ 
are potentially large because the fixed point value of 
the messenger coupling $\lambda_{irs}$
in Eq.~(\ref{messenger}) is not necessarily small}\/.
In particular,
slepton mass $m_{\tilde{e}}^2\fun{M_C}$
will be affected by a correction of order $\alp{3}M_3^2$ 
if the right-handed lepton $e_\R^c$ couples 
to colored SC fields through the messenger interactions.
Nevertheless, the presence or absence of such corrections
depends on the structure of the SC sector; for instance,
there is no two-loop term of $\lambda^2_{e\,}\alp{3}M_3^2$
if {\it both}\/ of $\Phi_r$ and $\Phi_s$ are $SU(3)$ singlets
in the messenger interaction $e_{\R\,}^c\Phi_r\Phi_s$.
In the following analysis,
we will assume that such higher-loop terms are negligible 
or at most comparable to the one-loop term\rlap.\,\footnote{
In the MSSM case with GUT relation of gaugino masses,
we do not expect that the presence of the $\alp{3}M_3^2$ term 
significantly changes the previous estimation of 
$\Delta_{\tilde{L}}$ and $\Delta_{\tilde{e}}$,
unless it is associated with a large group-theoretical factor.
}
%With this assumption, we may well justify the above formulas
%for sfermion masses at the decoupling scale $M_C$.
Another remark is that
there might arise large threshold effects
when the strongly-coupled SC sector decouples.
In the expression (\ref{diff-12}),
we have implicitly assumed that 
these are negligible or flavor independent.
We shall comment on such effects
in section~\ref{sec:threshold}.

In passing, we note the sparticle mass spectrum at the weak scale.
As far as the first two families of sfermions are concerned,
the convergent values at $M_C$ are quite small, 
$m^2_{\tilde f i} \approx 0$,
and the mass spectrum is similar to that in `no-scale' scenario,
provided that there is no large correction at the SC threshold.
{}For $M_C =10^{16}$~GeV, we have 
\begin{equation}
\left(m_{\tilde{Q}},\,m_{\tilde{L}},\,m_{\tilde{e}}\right)
= \left(0.91,\,0.25,\,0.13\right)M_3 \ .
\label{mass-spectrum}
\end{equation}
The other $SU(2)$-singlet squarks have masses similar to 
$m_{\tilde Q}$.
Again we have taken ${\cal S}=0$ for simplicity
although nonzero ${\cal S}$ is helpful to avoid
charged lightest superparticle (LSP)~\cite{KT}.
On the other hand, the third family, in particular the top quark, 
as well as Higgs fields do not couple to the SC sector. 
In general, their soft scalar masses
depend on initial values as well as the details of 
the RG running of gauge and Yukawa couplings.
Hence we restrict ourselves to the first and second 
families of sfermion massses in the following analysis.

\subsection{Mixing angles and FCNC constraints}
\label{subsec:FCNC}
\setcounter{footnote}{0}

The factor $\Delta_{\tilde f}$ represents 
a simple estimate of sfermion mass difference in the sfermion basis.
To confront the NS scenario
with the exprimental bounds on FCNC processes~\cite{FCNC},
we still have to evaluate sfermion mass matrices
in a basis that diagonalizes fermion Yukawa matrices;
specifically we are interested in the mixings
$(\delta_{12})_{\L\L,\,\R\R}$ between the first two families 
of left-handed, or right-handed sfermions.
In this respect,
the NS scenario has an advantage that
both of fermion and sfermion mass matrices are known
if the anomalous dimensions are specified.
If the first two families have the same anomalous dimension 
$\gamma_{f1}^*=\gamma_{f2}^*$ at the SC fixed point,
Yukawa-diagonalizing angle will be of $\order{1}$,
but at the same time, $\Gamma_{\tilde{f}1}=\Gamma_{\tilde{f}2}$ 
guarantees the complete mass degeneracy of sfermions.
On the other hand, if $\gamma_{f1}^* > \gamma_{f2}^*$ and 
$\Gamma_{\tilde f 1} > \Gamma_{\tilde f 2}$, 
the sfermions have non-degeneracy estimated by 
$\left(\Gamma_{\tilde{f}2}^{-1}-\Gamma_{\tilde{f}1}^{-1}\right)
\Delta_{\tilde f}$,
but the diagonalizing angle is as small as 
$(M_C/M_0)^{\gamma_1^*-\gamma_2^*}$, 
which gives an additional suppression to FCNC processes.
In this way, we expect that approximate alignment happens.

{}For squarks, the degeneracy factors are fairly small 
already in the sfermion basis. Moreover,
if the diagonalizing angles are of the order of the Cabbibo angle, 
we have $(\delta^d_{12})_{\L\L,\,\R\R} 
\sim 0.22 \times \Delta_{\tilde f} \approx 1 \times 10^{-3}$.
Using the constraint from the $K^0$--$\bar{K}^0$ system, \ie,
\begin{eqnarray}
\left(\delta^d_{12}\right)_{\L\L,\,\R\R} 
 < 1.2 \times 10^{-3} \times 
   \left(\frac{m_{\tilde d}}{500\,{\rm GeV}}\right) \ ,
\end{eqnarray}
it is required that 
$m_{\tilde d} \gsim 500$~GeV,
which corresponds to $M_3 \gsim 500$~GeV
as well as $m_{\tilde e} \gsim 50$~GeV.
Thus FCNC constraints will easily be satisfied in squark secor.

On the other hand, 
the degeneracy factors are not small enough in slepton sector.
Then lepton flavor-violating processes such as 
$\mu \rightarrow e \gamma$ decay constrain the slepton masses.
Of course, such constrains depend on the texture of 
lepton Yukawa matrix generated by the NS mechanism.
To see this explicitly, let us concentrate on the first two
families and denote by $\theta_{\L}$ and $\theta_\R$ 
the mixing angles for left- and right-handed leptons, respectively.
{}First consider the case in which 
left-handed leptons receive the same anomalous dimension 
$\gamma_{L1} = \gamma_{L2}$ from the SC dynamics.
Then the first $2 \times 2$ lepton Yukawa matrix takes, 
up to $\order{1}$ factors, the `lopsided' form 
\begin{eqnarray}
y_{ij}\,\approx\,
y_{22} \left(\matrix{
       \varepsilon & 1 \cr
       \varepsilon & 1    }\right) \ , \qquad
\varepsilon
 = \left(\frac{M_C}{M_0}\right)^{\gamma_{e1}^*-\gamma_{e2}^*}
\,\sim\,\frac{m_e}{m_\mu}\,=\,5 \times 10^{-3} \ ,
\end{eqnarray}
where the lepton mass hierarchy originates from 
the anomalous dimensions of the right-handed leptons.
In this case, the left-handed leptons have $\order{1}$ mixing and 
that will be favorable from the viewpoint of neutrino oscillation.
Interestingly, the left-handed sleptons are completely degenerate
since $\gamma_{L1}^*=\gamma_{L2}^*$ implies $\Gamma_{L1}=\Gamma_{L2}$.
Moreover, the right-handed leptons have the mixing angle of 
$\theta_\R=\order{\varepsilon}$,
which gives an additional suppression to $(\delta_{12}^\ell)_{\R\R}$.
Thus we will have no strong constraint.

If mixing angles are larger than $\order{10^{-3}}$, 
the constraint becomes significant.
{}For example, if both of mixing angles 
for left- and right-handed sectors are of the order
\begin{eqnarray}
\theta_\L\,\sim\,
\theta_\R\,\sim\,\sqrt{\frac{m_e}{m_\mu}}
\,=\,7 \times 10^{-2} \ ,
\label{mixing:both}
\end{eqnarray}
we have $(\delta^\ell_{12})_{\L\L}\sim 7\times10^{-2} 
\Delta_{\tilde f} \approx 3 \times 10^{-3}$ 
and $(\delta^\ell_{12})_{\R\R}\sim 7\times10^{-2}  
\Delta_{\tilde f} \approx 7 \times 10^{-3}$, and 
it is required that $m_{\tilde L} \gsim 100$~GeV and 
$m_{\tilde e} \gsim 200$~GeV from the constraint 
\begin{eqnarray}
\left(\delta^\ell_{12}\right)_{\L\L,\,\R\R}
 < 2.0 \times 10^{-3}\times
   \left(\frac{m_{\tilde \ell}}{100\,{\rm GeV}}\right)^2 \ .
\end{eqnarray}
The constraint on $(\delta^\ell_{12})_{\R\R}$ requires that
the gluino and squarks are heavier than $2\,{\rm TeV}$, 
whereas that on $(\delta^\ell_{12})_{\L\L}$ corresponds to
the gluino and squarks heavier than $400$~GeV.
In general, the larger mixing angle 
the right-handed lepton sector has, the heavier sleptons 
are required from $\mu \rightarrow e \gamma$ decay.
{}For example, it is required that
$m_{\tilde{e}}\gsim 300$~GeV if $\theta_\R\gsim{}0.2$.

We remark that there are additional constraints 
coming from $(\delta_{12})_{\L\R}$,
which are related to SUSY-breaking trilinear couplings 
$h_{ij\,}\wt{\psi}_i\wt{\psi}_j{}H$.
In the present SCFT framework,
soft scalar masses are well controlled by the SC dynamics,
but the so-called $A$-terms $A_{ij} \equiv h_{ij}/y_{ij}$ are not; 
the SC dynamics does give a suppression factor to $h_{ij}$, 
which is the same as the suppression factor of the corresponding 
Yukawa coupling $y_{ij}$.
Therefore, even if the constraints on 
$\left(\delta_{12}\right)_{\L\L,\,\R\R}$ are satisfied, 
we should still take care of additional contraints
on $(\delta_{12})_{\L\R}$ from FCNC processes~\cite{NS2}.
We need another mechanism to suppress $A$-terms,
or to realize the complete alignment\footnote{
One way to achieve this possibility
was discussed in Ref.~\cite{KY},
where the Yukawa couplings $y_{ij}$ by themselves 
are suppossed to have infrared fixed points.
In models with extra dimensions,
it was also argued~\cite{BKNY} that  
thanks to power-law behavior due to Kaluza-Klein modes,
hierarchical Yukawa couplings can be obtained at the IR fixed points.
In this case, however, it seems difficult to realize 
finite mixing angles. 
}\,of $A_{ij}$. At any rate, we expect that
the constraints coming from $(\delta_{12})_{\L\R}$ will be 
less severe than those from $(\delta_{12})_{\L\L,\,\R\R}$,
and to examine the latter is the subject of the following sections.

%%%%%%%%%%%%%%%%%%%%%%%%%%%%%%%%%%%%%%%%%%%%%%%%%%%%%%%%%%%%%%%%%%%%%%
\section{$\bm{SU(5)}$ GUT coupled to SC sector}
\label{sec:GG}
%%%%%%%%%%%%%%%%%%%%%%%%%%%%%%%%%%%%%%%%%%%%%%%%%%%%%%%%%%%%%%%%%%%%%%

%The degeneracy factor $\Delta_{\tilde f}$ for the right-handed 
%sleptons is not small enough for the MSSM coupled to SC sectors.
%The main reason is that 
%right-handed sleptons carry only the $U(1)_Y$ charge 
%and receive small radiative correction in the `MSSM regime', 
%\ie, in the energy scale between $M_C$ and $M_Z$.
%{}For a given mass difference (\ref{diff-12}),
%the smaller the radiative correction is, the worse the degeneracy is.
%[Generically, the degeneracy factor slowly becomes small 
%as we take $M_C$ large, 
%since the radiative correction increases logarithmically.]

For $M_C > 10^{16}$~GeV, we are naturally led to 
consider the embedding of the MSSM into SUSY GUTs.
In this section, we study the Georgi-Glashow $SU(5)$ model, 
which breaks down to the MSSM 
at the scale $M_X = 2 \times 10^{16}$~GeV.
We assume $M_C\ge{}M_X$, \ie, the SC sector decouples 
at a larger scale than the GUT breaking scale.

As in the MSSM case of subsection~\ref{subsec:mass}, 
we expect that each sfermion mass converges on
\begin{eqnarray}
  m^2_{\tilde f i}\fun{M_C}
 = \frac{1}{\Gamma_{\tilde f i}}\,
   C_{f 5}\,\alp{5}\fun{M_C}M_5^2(M_C) \ ,
\label{converge:su5}
\end{eqnarray}
where $\alpha_5(M_C)$ and $M_5(M_C)$ are 
the $SU(5)$ gauge coupling and gaugino mass at the scale $M_C$.
Representations of quarks and leptons under the $SU(5)_{{\rm GG}}$ 
are shown in the fifth column of Table~\ref{table:coefficient}.
The sixth column shows $C_{f 5}$ for each matter field.
On the other hand, the radiative correction 
in the `GUT regime' between $M_C$ and $M_X$ is obtained as
\begin{eqnarray}
\label{gutmassdiff}
\Delta m^2_{\tilde f i}(M_C \rightarrow M_X)
 = \frac{C_{f 5}}{2b_5}\,I_5\fun{M_X,M_C}M_5^2(M_C) \ ,
\end{eqnarray}
where $b_5$ is the beta-function coefficient of $\alpha_5$, 
and $I_5(M_X,M_C)$ is defined by
\begin{equation}
I_5(M_X,M_C) \equiv  1 - \frac{\alpha_5^2(M_X)}{\alpha^2_5(M_C)} \ .
\end{equation}
Then the sfermion mass degeneracy may be estimated by the use of 
\begin{equation}
\Delta_{\tilde f }
 = \frac{C_{f 5}\,\alp{5}\fun{M_C}M_5^2(M_C)}
        {\Delta m^2_{\tilde f i}(M_C \rightarrow M_X) 
        +\Delta m^2_{\tilde f i}(M_X \rightarrow M_Z)} \ ,
\end{equation}
where $\Delta m^2_{\tilde f i}(M_X \rightarrow M_Z)$ 
denotes the radiative correction in the MSSM regime,
Eq.~(\ref{rad-cor}) with $M_C = M_X$.
The value of $\Delta_{\tilde f }$ depends on the flow of 
the gauge couplings, $\alpha_5$ and $\alpha_{a=1,2,3}$,
but not the gaugino masses.

\begin{figure}[tb]
\vspace*{-2.5cm}
\begin{center}
\epsfxsize=0.6\textwidth
\leavevmode
\epsffile{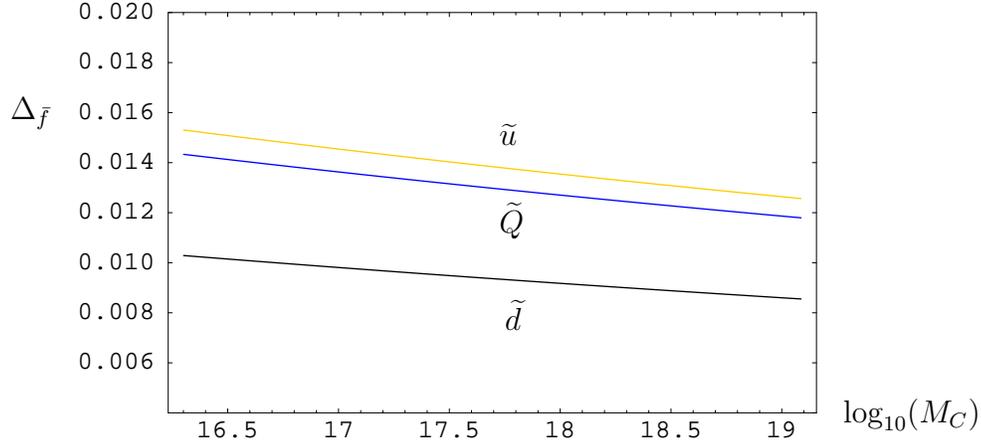}
\put(-305,180){$\Delta_{\bar f}$}
\put(-120,170){$\widetilde u$}
\put(-120,137){$\widetilde Q$}
\put(-118,100){$\widetilde d$}
\put(10,65){$\log_{10}(M_C)$}\\[-2cm]
\caption{$\Delta_{\widetilde Q}$\,, $\Delta_{\widetilde u}$ 
and $\Delta_{\widetilde d}$
  against $M_C$ in the supersymmetric $SU(5)$ GUT.}
\label{fig:su5Qud}
\end{center}
\end{figure}

\begin{figure}[htb]
\vspace*{-1.5cm}
\begin{center}
\epsfxsize=0.56\textwidth
\leavevmode
\hspace{6mm}
\epsffile{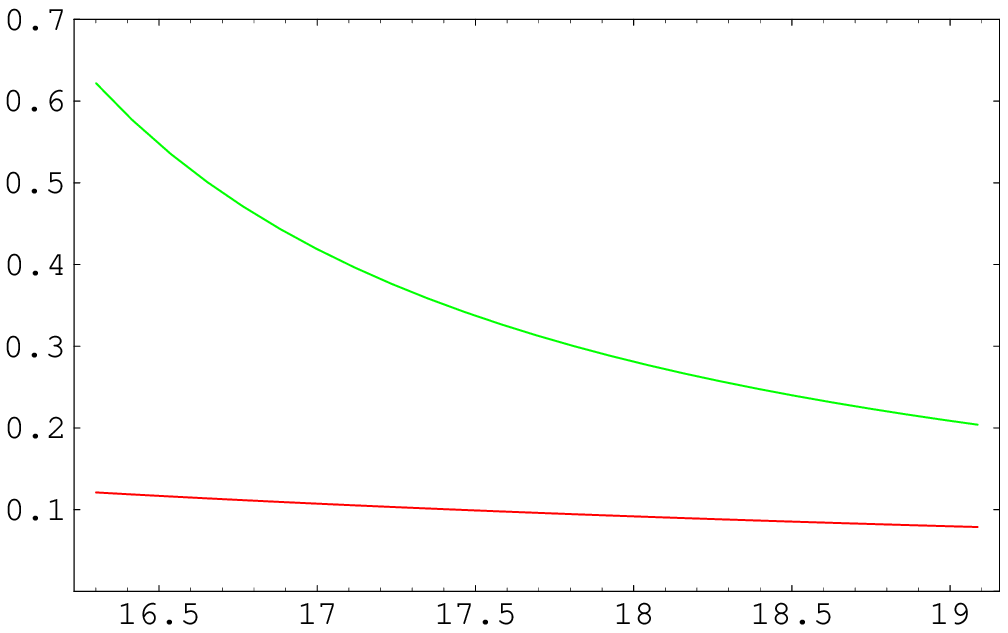}
\put(-300,160){$\Delta_{\bar f}$}
\put(-120,135){$\widetilde e$}
\put(-121,85){$\widetilde L$}
\put(10,57){$\log_{10}(M_C)$}\\[-1.7cm]
\caption{$\Delta_{\widetilde e}$ and $\Delta_{\widetilde L}$
against $M_C$ in the supersymmetric $SU(5)$ GUT.}
\label{fig:su5Le}
\end{center}
\end{figure}

Fig.~\ref{fig:su5Qud} shows $\Delta_{\tilde f}$ for squarks and 
Fig.~\ref{fig:su5Le} for sleptons.
{}For definiteness, we have taken $b_5 = -9/2$
corresponding to the minimal $SU(5)$, 
while different $b_5$ lead to similar results.

Compared with the MSSM case, 
Eqs.~(\ref{Delta:MSSM})--(\ref{mass-spectrum}),
the mass spectrum at the weak scale is almost the same, 
but the degeneracy factors are not.
{}First of all, the degeneracy factors 
for squarks and left-handed slepton become slightly large.
Note also that
$\Delta_{\tilde Q}$ and $\Delta_{\tilde u}$ are larger 
than $\Delta_{\tilde{d}\,}$, since $Q$ and $u$ belong to 
the $\bm{10}$ representation under the $SU(5)$
whereas $d$ belongs to the $\wb{\bm{5}}$, 
and the former has a larger convergent value than the latter.
There is a small splitting between 
$\Delta_{\tilde Q}$ and $\Delta_{\tilde u}$ because
$m_{\tilde Q}^2$ gets slightly larger radiative corrections 
in the MSSM regime from the wino mass $M_2$.

More importantly, the degeneracy for the right-handed slepton 
becomes drastically worse than before.
The reason is very simple; 
the right-handed lepton is embedded into the $\bm{10}$, 
which has a large Casimir coefficient, 
so that the convergent value (\ref{converge:su5}) is enlarged.
Consequently the FCNC constraint for $(\delta^\ell_{12})_{\R\R}$ 
cannot easily be satisfied, requiring heavier sleptons 
and/or a specific form of lepton Yukawa matrix.
It is rather surprising that
the simple group-theoretical fact precludes
the $SU(5)$ extension of the MSSM. At any rate, 
the embedding into the $SU(5)$ discussed here would remove
one of most interesting properties of the present SCFT approach, 
that is, approximate mass degeneracy of sfermions 
without specifying a mediation mechanism of SUSY breaking.

%%%%%%%%%%%%%%%%%%%%%%%%%%%%%%%%%%%%%%%%%%%%%%%%%%%%%%%%%%%%%%%%%%%%%%
\section{Flipped $\bm{SU(5) \times U(1)}$ coupled to SC sector}
\label{sec:flipped}
\setcounter{footnote}{0}
%%%%%%%%%%%%%%%%%%%%%%%%%%%%%%%%%%%%%%%%%%%%%%%%%%%%%%%%%%%%%%%%%%%%%%

We have observed that 
within the framework of the ordinary $SU(5)$ GUT,
the degeneracy factor $\Delta_{\tilde e}$ 
is not small enough for the right-handed slepton
because of its large Casimir coefficient.
{}From this viewpoint, it is interesting to consider
the flipped $SU(5)\times U(1)$ models~\cite{Flip1}, 
where the right-handed lepton is not embedded into 
a nontrivial $SU(5)$ representation, but is an $SU(5)$-singlet $S_i$.

Let us first recall the basic feature of the minimal 
flipped $SU(5)_\F\times{}U(1)_\F$ model to fix our notation.
The quantum number of matter and Higgs fields
is shown in Table~\ref{table:flipped}, and
the embedding of each family of matter fields 
is explicitly given by 
\begin{eqnarray}
T= \pmatrix{0 & d^c_{\R 3} &-d^c_{\R 2} & u_1 & d_1      \cr
              & 0          & d^c_{\R 1} & u_2 & d_2      \cr
              &            & 0          & u_3 & d_3      \cr
              &            &            & 0   & \nu_\R^c \cr
              &            &            &     & 0           
            } \,, \qquad
\wb{F}
 = \pmatrix{u^c_{\R 1} \cr u^c_{\R 2} \cr u^c_{\R 3} \cr -e \cr \nu
            } \,, \qquad
S= e^c_\R \ ,
\end{eqnarray}
where $\nu_\R^c$ is a right-handed neutrino.
\begin{table}[htbp]
\begin{center}
\begin{tabular}{|c||c|c|c|}
\hline
\makebox[10mm]{%Superfields
               } & 
\makebox[30mm]{MSSM content} &
\makebox[25mm]{$SU(5)_\F$ rep.} &
\makebox[25mm]{$U(1)_\F$ charge}\\
\hline\hline
$T_i$      & $d_i,\,Q_i,\,\nu^c_{\R i}$ 
	   & $\bm{10}$      & $+\,1$    \\ \hline
$\wb{F}_i$ & $u_i,\ L_i$ 
           & $\wb{\bm{5}}$  & $-\,3$    \\ \hline
$S_i$      & $e^c_{\R i}$ 
           & $\bm{1}$       & $+\,5$    \\ \hline
$h$        & $h^c_d,\ h_d^{\phantom{c}}$ 
           & $\bm{5}$       & $-\,2$    \\ \hline
$\wb{h}$   & $h^c_u,\ h_u^{\phantom{c}}$ 
           & $\wb{\bm{5}}$  & $+\,2$    \\ \hline
$H$        & $\bm{-\!\!\!-\!\!\!-}$
           & $\bm{10}$      & $+\,1$    \\ \hline
$\wb{H}$   & $\bm{-\!\!\!-\!\!\!-}$
	   & $\wb{\bm{10}}$ & $-\,1$    \\ \hline
\end{tabular}
\caption{The minimal flipped $SU(5)_\F\times U(1)_\F$ model:
The last column shows $F\equiv\sqrt{40}\,\wt{X}$, 
where $\wt{X}$ is the $SO(10)$-normalized $U(1)_\F$ charge
of Eq.~(\ref{generator:flipped}).
}
\label{table:flipped}
\end{center}
\end{table}
If the $\nu_\R^c$ components of the $10$-dimensional Higgs fields 
$H$ and $\wb{H}$ develop the vacuum expectation values,
the symmetry breaking to the MSSM can be achieved
with a simple implimentation of the missing partnar mechanism 
for doublet-triplet splitting \cite{missingpartnar}.
Through this symmetry breaking, the $24$-th generator 
$T^{24}_\F\equiv\sqrt{3/5}\,\wt{Y}$ of $SU(5)_\F$ and 
the $U(1)_\F$ generator $F\equiv\sqrt{40}\,\wt{X}$ are related to 
the hypercharge $Y$ as well as its orthogonal broken generator $X$.
One way to describe this relation is 
to embed the gauge group into $SO(10)$ and 
to represent $U(1)$ generators in the $SO(10)$ basis.
By picking its $SU(4)_\PS\times{}SU(2)_\L\times{}SU(2)_\R$ subgroup,
we have
\def\lsize#1{\hbox{\Large ${#1}$}}
\begin{eqnarray}
\wt{X}
 \,=\,\lsize{\sqrt{\frac{3}{5}}}\,T_\PS^{15}
     +\lsize{\sqrt{\frac{2}{5}}}\,T_\R^3 \ , \qquad
\lsize{\sqrt{\frac{3}{5}}}\,\wt{Y}
 \,=\,\lsize{\sqrt{\frac{2}{5}}}\,T_\PS^{15}
     -\lsize{\sqrt{\frac{3}{5}}}\,T_\R^3 \ .
\label{generator:flipped}
\end{eqnarray}
Here $T_\R^3$ is the third generator of $SU(2)_\R$ 
and $T_{{\rm PS}}^{15}=\sqrt{3/8}\left(B-L\right)$ 
is the $15$-th generator of the Pati-Salam $SU(4)$, 
which is proportional to the $B-L$ charge.
%Note that $T_\R^3=-1/2$ for $u_\R^c$ and $\nu_\R^c$,
%and $T_\R^3=+1/2$ for $d_\R^c$ and $e_\R^c$.
Eq.~(\ref{generator:flipped}) is precisely 
the flipped version of the usual $SO(10)$ relation
\begin{eqnarray}
X\,=\,\lsize{\sqrt{\frac{3}{5}}}\,T_\PS^{15}
     -\lsize{\sqrt{\frac{2}{5}}}\,T_\R^3 \ , \qquad
\lsize{\sqrt{\frac{3}{5}}}\,Y
 \,=\,\lsize{\sqrt{\frac{2}{5}}}\,T_\PS^{15}
     +\lsize{\sqrt{\frac{3}{5}}}\,T_\R^3 \ ,
\label{generator:unflipped}
\end{eqnarray}
where $X$ is nothing but the $U(1)$ generator 
of $SO(10)/SU(5)_{{\rm GG}}$, often called $U(1)_\chi$.
Since these two sets of $U(1)$ generators are 
orthogonally related to each other, we have
\begin{eqnarray}
X\,=\,\frac{1}{5}\,\wt{X}
     +\frac{\sqrt{24}}{5}\,T_\F^{24} \ , \qquad
\lsize{\sqrt{\frac{3}{5}}}\,
Y\,=\,\frac{\sqrt{24}}{5}\,\wt{X}
     -\frac{1}{5}\,T_\F^{24} \ .
\label{hypercharge}
\end{eqnarray}
{}For notational simplicity,
let $\alphafive{}$ and $\alphaone{}$ denote
the gauge couplings of $SU(5)_\F\times{}U(1)_\F$.
With the normalization of $U(1)$ generators as above,
the embedding into the $SO(10)$ GUT 
would lead to $\alphaone{} = \alphafive{}$,
although we do not assume such a further unification.

We assume that the SC sector decouples at the scale $M_C$,
below which the model is exactly the same as 
the minimal flipped $SU(5)\times U(1)$ model.
At the scale $M_X$ where 
$SU(5)\times{}U(1)$ breaks to the MSSM gauge group,
the hypercharge gauge coupling $\alpha_Y=\left(3/5\right)\alpha_1$
is matched with $SU(5)\times{}U(1)$ gauge couplings 
$\alphafive{}$ and $\alphaone{}$ 
according to
\begin{eqnarray}
\frac{15}{\alpha_Y}\,\equiv\,\frac{25}{\alpha_1}
\,=\,\frac{1}{\alphafive{}}+\frac{24}{\alphaone{}} \ .
\label{matching:gauge}
\end{eqnarray}
The measured value of the SM gauge couplings requires
$\alphaone{}\fun{M_X}\approx\alphafive{}\fun{M_X}
\approx\left(2\pi\times{}24.5\right)^{-1}$.
The matching condition for the corresponding gaugino masses 
is given by
\begin{eqnarray}
\frac{M_1}{\alpha_1}
\,=\,\frac{1}{25}
     \left(\frac{\Mfive{}}{\alphafive{}}
          +\frac{24\Mone{}}{\alphaone{}}\right)
  =  \left(\frac{1+24\ratio}{25}\right)
     \frac{\Mfive{}}{\alphafive{}}
  =  \left(\frac{1+24\ratio}{25\ratio}\right)
     \frac{\Mone{}}{\alphaone{}} \ .
\label{matching:gaugino}
\end{eqnarray}
%as is seen in the spurion formalism from the fact that
%the gaugino mass appears through 
%the combination $\alpha_a^{-1}\left(1-M_a\theta^2\right)$.
Here we have introduced an RG-invariant quantity
\begin{equation}
\ratio\,\equiv\,
     \frac{\Mone{\,}/\alphaone{}}{\Mfive{\,}/\alphafive{}}
\,\approx\,\frac{\Mone{}\fun{M_X}}{\Mfive{}\fun{M_X}} \ ,
\label{def:r}
\end{equation}
which parametrizes the relative size of 
two independent gaugino masses of $SU(5)\times{}U(1)$.
$SO(10)$ gauge symmetry would require $\ratio=+1$,
but we treat $\ratio$ as a free parameter.
The gluino mass $M_3$ and the wino mass $M_2$ 
are matched as usual and are independent of $\Mone{}$.
However, the bino mass $M_1$ does depend on $\ratio$,
and the ratio $M_1/M_3$ is given by
\begin{equation}
{M_1 \over M_3}
\,=  \left(\frac{1+24\ratio}{25}\right)\frac{\alpha_1}{\alpha_3}
\,=  \left(\frac{1+24\ratio}{25}\right)\frac{5\alpha_Y}{3\alpha_3} \ .
\label{gaugino:r}
\end{equation}
We expect that the degeneracy factor 
for the right-handed sleptons also depend on $\ratio$, since 
their masses will be dominantly governed by the $U(1)$ gaugino masses.

It is straighforward to estimate sfermion masses
and confirm our expectation.
The RGE for soft scalar mass, 
which includes the leading term of 
$SU(5)\times{}U(1)$ gauge loop, is
%\footnote{
%As mentioned in subsection~\ref{subsec:mass},
%the second term in RHS of Eq.~(\ref{rg-scalar-m-FP}) 
%is the one-loop contribution 
%due to the $SU(5) \times U(1)$ gaugino masses, 
%whereas the first term includes full effects from the SC sector.
%Here we assume that the SC-sector fields that couple to $e_\R^c$
%are $SU(5)$-singlets; otherwise,
%the term proportional to $\alphafive{}\Mfive{2}$ 
%would appear in the RGE at two-loop level 
%and the convergent value for the right-handed slepton mass 
%would be affected by $\alphafive{}\Mfive{2}$ as well.}
\begin{equation}
\mu\frac{dm^2_{i}}{d\mu}
 = {\cal M}_{ij\,}m^2_{j} 
  -\sum_{a=5,\flipped}\CFlipped{ia}\,
   \alphaFlipped{a}{}\MFlipped{a}{2} \ ,
\label{rg-scalar-m-FP}
\end{equation}
where $\CFlipped{ia}$
%$\CFlipped{ia}=4C_2\fun{R_{ia}}$ ($a=5,\,\flipped$) 
denote the Casimir factors under $SU(5)\times{}U(1)$.
Under the assumption that
higher order corrections to the second term is small,
we may estimate the convergent value of each sfermion mass
at $M_C$ and the degeneracy factor at the weak scale by
\begin{eqnarray}
  m^2_{\tilde f i}\fun{M_C}
&=&\frac{1}{\Gamma_{\tilde f i}}\sum_{a=5,\flipped}
   \CFlipped{fa}\,\alphaFlipped{a}{}\fun{M_C}\MFlipped{a}{2}(M_C) \ ,
\label{convergence:flipped}\\
\Delta_{\tilde f }
&=&\frac{\sum_{a=5,\flipped}
         \CFlipped{fa\,}\alphaFlipped{a}{}\fun{M_C}\MFlipped{a}{2}(M_C)}
        {\Delta m^2_{\tilde f i}(M_C\!\rightarrow\!M_X)
        +\Delta m^2_{\tilde f i}(M_X\!\rightarrow\!M_Z) } \ .
\end{eqnarray}
The group-theoretical factors are shown 
in Table~\ref{table:coefficient2},
and the radiative corrections in $SU(5)\times{}U(1)$ regime
can be calculated in a manner similar to Eq.~(\ref{mssmmassdiff})
with $\bfive=-5$ and $\bone=15/2$.
The results does not strongly depend on $M_C$,
and we will take $M_C=M_X$ for definiteness.

\begin{table}[tb]
  \begin{center}
    \leavevmode
    \begin{tabular}[t]{|c||c|c|c|c|c|}
      \hline
\makebox[10mm]{$f$} & 
\makebox[30mm]{$SU(5)\times U(1)$} & 
\makebox[18mm]{$\Cfive{f}$} & 
\makebox[18mm]{$\Cone{f}$} &
\makebox[15mm]{$6\wt{Y}$} &
\makebox[15mm]{$\sqrt{40}\,X$} \\
\hline\hline
$Q$ & $(\bm{10},+1)$ & $72/5$ & $1/10$ & $+1$ & $+1$\\
\hline
$d_\R^c$ & $(\bm{10},+1)$ & $72/5$ & $1/10$  & $-4$ & $-3$\\
\hline
$\nu_\R^c$ & $(\bm{10},+1)$ & $72/5$ & $1/10$ & $+6$ & $+5$\\
\hline
$L$ & $(\wb{\bm{5}},-3)$ & $48/5$ & $9/10$ & $-3$ & $-3$\\
\hline
$u_\R^c$ & $(\wb{\bm{5}},-3)$ & $48/5$ & $9/10$ & $+2$ & $+1$\\
\hline
$e_\R^c$ & $(\bm{1},+5)$ & $0$ & $5/2$ & $\ \ 0$ & $+1$\\
\hline
\end{tabular}
\caption{
The group-theoretical factors
in the flipped $SU(5)\times{}U(1)$ model:
the last two column shows the `flipped hypercharge' $\wt{Y}$
and the $U(1)_\chi$ charge $X$
in Eqs.~(\ref{generator:flipped})--(\ref{generator:unflipped}).
}
\label{table:coefficient2}
\end{center}
\end{table}

\begin{figure}[tb]
\vspace*{-3.5cm}
\begin{center}
\epsfxsize=0.5\textwidth
\leavevmode
\epsffile{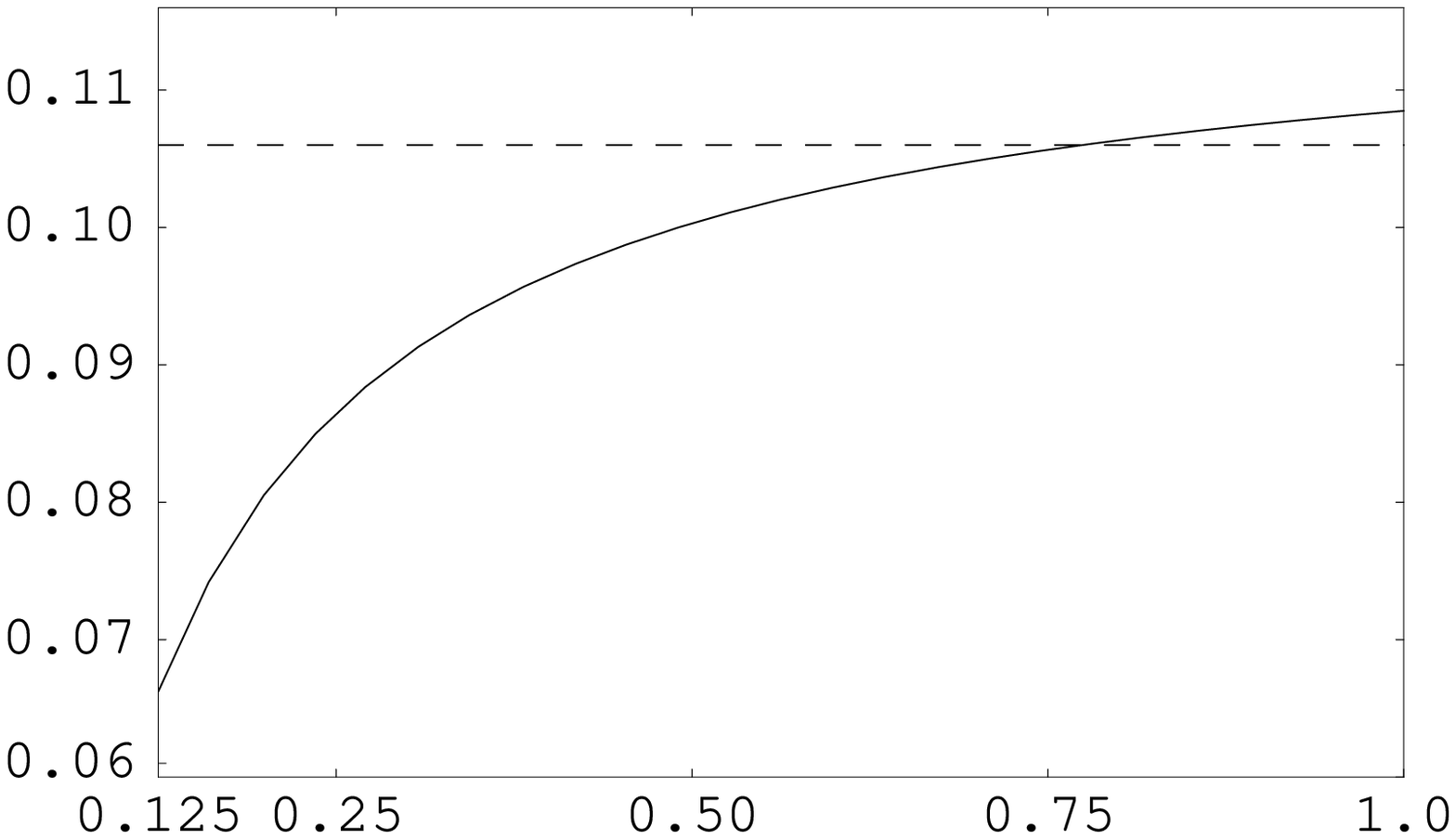}
\put(-260,200){$\Delta_{\tilde e}$}
\put(10,98){$\ratio$}\\[-3.5cm]
\caption{The slepton degeneracy factor $\Delta_{\widetilde e}$ 
against $\ratio$ in the flipped $SU(5) \times U(1)$ model.
The dotted line shows the value in the MSSM case.
}
\label{fig:Delta_e}
\end{center}
\end{figure}

We are especially interested in the expected degree of 
the degeneracy $\Delta_{\tilde{e}}$ for the right-handed sleptons. 
The point is that 
the convergent value $m^2_{\tilde{e}}\fun{M_C}$
is governed by the flipped $U(1)$ gaugino mass $\Mone{}$,
whereas the radiative corrections $\Delta{}m^2_{\tilde{e}}$ 
are determined by the bino mass $M_1$, 
which contains as a component the $SU(5)$ gaugino mass $\Mfive{}$
as well as $\Mone{}$
according to the matching condition~(\ref{matching:gaugino}).
Consequently, the degeneracy factor $\Delta_{\tilde e}$ behaves like
\begin{eqnarray}
\Delta_{\tilde{e}}
\ \propto\ \left(\frac{\Mone{}}{M_1}\right)^2
\ \propto\ \left(\frac{25\ratio}{1+24\ratio}\right)^2 \ .
\end{eqnarray}
Numerical values for $\Delta_{\tilde e}$ 
are shown in Fig.~\ref{fig:Delta_e} as a function of $\ratio$,
and we find explicitly 
\begin{eqnarray}
\Delta_{\tilde e}\,
 = \left\{\begin{array}{lcl}
   6.6 \times 10^{-2} &\hbox{for} & \ratio=1/8 \\
   1.1 \times 10^{-1} &\hbox{for} & \ratio=1 
   \end{array}\right.\ .
\end{eqnarray}
Compared with the MSSM case (\ref{Delta:MSSM}),
a similar value of $\Delta_{\tilde{e}}$
is obtained in the $\ratio=1$ case,
as is expected from the fact that $\ratio\approx1$ 
corresponds to the usual GUT relation of gaugino masses.
As we decrease $\ratio$ with keeping $\Mfive{}$ fixed,
$\Mone{}$ decreases linearly in $\ratio$,
whereas $M_1$ reaches a nonzero value 
and thus $\Delta_{\tilde e}$ becomes small.
In other words,
{\it the mass difference of $\wt{e}_\R^c$ is determined 
by a small gaugino mass of the flipped $U(1)$}\/
while {\it the $SU(5)$ component of the bino mass
effectively enhances the averaged slepton mass}\/.
In this way, 
we find that the degeneracy factor $\Delta_{\tilde{e}}$ is 
improved for $\ratio\lsim 0.5$ compared with the MSSM case.
{}For $\abs{\ratio}\gg{}1$,
the bino mass $M_1$ becomes proportional to $\Mone{}$,
and $\Delta_{\tilde e}$ approaches its maximal $0.12$.

At first sight, the degeneracy factor $\Delta_{\tilde e}$
could be arbitrarily small if $\Mone{}\ll\Mfive{}$.
However, this is not true because 
our estimation based in the RGE (\ref{rg-scalar-m-FP}) 
is no longer reliable for a hierarchically small value
of $\abs{\ratio}\lsim0.1$.
At two-loop level, for instance, 
the RGE contains potentially dangeous terms like 
$\lambda^2_{e\,}\alphafive{}\Mfive{2}$ and 
$\alphaone{}\alphafive{}\Mone{}\Mfive{}$, where
$\lambda_e$ is the messenger coupling of the $SU(5)$-singlet $e_\R^c$.
As mentioned in subsection~\ref{subsec:mass},
the former term is absent if we assume that only $SU(5)$-singlet
fields couple to $e_\R^c$ through the messenger interaction.
On the other hand, 
the latter term will give a substantial correction to 
$\Delta_{\tilde{e}}$ for such a small value of $\abs{\ratio}$.
Even if the latter correction is properly taken into account,
the ratio $M_1/M_3$ would be unacceptably small
from the viewpoint of the naturalness.

\begin{figure}[tb]
\vspace*{-3.5cm}
\begin{center}
\epsfxsize=0.5\textwidth
\leavevmode
\hspace{.1cm}
\epsffile{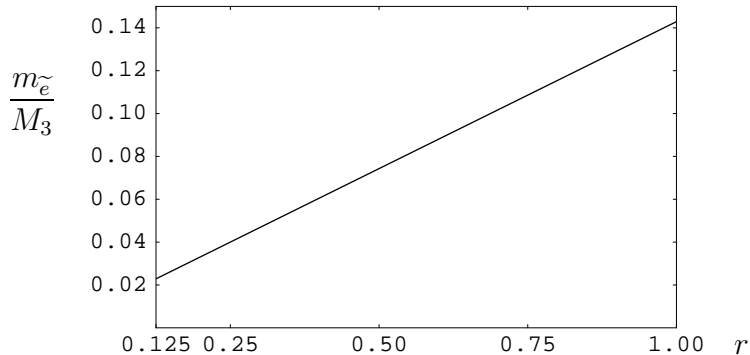}
\put(-265,190){${\displaystyle\frac{m_{\widetilde e}}{M_3}}$}
\put(10,97){$\ratio$}\\[-3.5cm]
\caption{$m_{\widetilde e}/M_3$ against $\ratio$ 
in the flipped $SU(5) \times U(1)$ model.}
\end{center}
\label{fig:m_e/M_3}
\end{figure}

%To be explicit, let us consider the slepton mass 
%instead of studying neutralino mass eigenvalues.
The slepton mass at the weak scale is approximately given by
\begin{equation}
m_{\tilde e} = 0.93 M_1 \ ,
\label{LSP}
\end{equation}
as long as the radiative correction $\Delta{}m^2_{\tilde{e}}$
dominates over the convergent value.
Eq.~(\ref{gaugino:r}) then implies that 
the ratio $m_{\tilde e}/M_3$ decrease as we decrease $\ratio$,
as is shown in Fig.~\ref{fig:m_e/M_3}.

{}For the other sfermions,
the degeneracy factors and mass spectrum 
are determined dominantly by $\alpha_a$ and $M_a$ ($a=2,3$),
and are almost independent of $\ratio$
for a moderate range of the parameter $\ratio\lsim\order{2.5}$.
Explicitly, we obtain for $\ratio=1/8$, 
\begin{eqnarray}
\Delta_{\tilde Q} &\!=\!& 1.5 \times 10^{-2} \ , \qquad 
\Delta_{\tilde L}  \,=\,  1.3 \times 10^{-1} \ , \cr
\Delta_{\tilde d} &\!=\!& 1.6 \times 10^{-2} \ , \qquad
\Delta_{\tilde u\,}\,=\,  1.1 \times 10^{-2} \ , 
\end{eqnarray}
which are almost the same as in the $SU(5)$ case,
except that $\tilde u$ and $\tilde d$ are interchanged.
The mass spectrum for squarks and left-handed sletpons is almost 
the same as in Eq.~(\ref{mass-spectrum}).

Noted that
$\Delta_{\tilde L}$ is larger than $\Delta_{\tilde e}\le 0.12$
owing to the embedding into $\wb{\bm{5}}$. 
As discussed in subsection~\ref{subsec:FCNC},
the constraints from lepton flavor violations
depend on the lepton mixing angles.
{}For mixing angles (\ref{mixing:both}),
the contraints from $(\delta^\ell_{12})_{\L\L}$ require
$m_{\tilde \ell} \gsim 200$~GeV, which should be compared with 
$m_{\tilde e} \gsim 200$~GeV from $(\delta^\ell_{12})_{\R\R}$ 
in the MSSM with the same mixing angles.
The former corresponds to $M_3\gsim 1$~TeV,
while the latter requires $M_3\gsim 2$~TeV.
Thus the flipped $SU(5)\times U(1)$ model, 
unlike the $SU(5)$ models, is on the level of the MSSM.

%%%%%%%%%%%%%%%%%%%%%%%%%%%%%%%%%%%%%%%%%%%%%%%%%%%%%%%%%%%%%%%%%%%%%%
\section{On threshold effects}
\label{sec:threshold}
%%%%%%%%%%%%%%%%%%%%%%%%%%%%%%%%%%%%%%%%%%%%%%%%%%%%%%%%%%%%%%%%%%%%%%

Up to now, we have smoothly connected RG flows 
of soft scalar masses at $M_{X\!}$ as well as $M_C$\/.
We reexamine this procedure and briefly discuss 
possible threshold effects of various sorts.

%%%%%%%%%%%%%%%%%%%%%%%%%%%%%%%%%%%%%%%%%%%%%%%%%%%%%%%%%%%%%%%%%%%%%%
\subsection{$\bm{D}$\,-\,term contributions in flipped models}
\label{subsec:Dterm}
%%%%%%%%%%%%%%%%%%%%%%%%%%%%%%%%%%%%%%%%%%%%%%%%%%%%%%%%%%%%%%%%%%%%%%

When the rank of gauge group is reduced, the gauge symmetry breaking 
induces the $D$-term contribution to soft scalar mass-squared, 
$\Delta_D m_{\tilde f}^2 = q_f\VEV{D}$, 
which is proportional to the charge of 
broken $U(1)$ symmetry~\cite{nonanomalous,anomalous}.
Here we consider effects of such additional terms at $M_X$ 
induced by $SU(5)\times U(1)$ breaking.
Of course, the charges under the broken symmetry are the same 
between different families, and those do not create non-degeneracy.
However, such additional contributions change 
overall magnitude of sfermion masses at the GUT threshold, 
and also change the mass spectrum and degeneracy factors 
at the weak scale.

The $D$-term contributions induced through the breaking 
of $SU(5)\times U(1)$ can be expressed in terms of 
the broken generators $F\equiv\sqrt{40}\,\wt{X}$ 
and $\wt{Y}\equiv\sqrt{5/3}\,T^{24}_\F$ 
in Eq.~(\ref{generator:flipped}) as
\begin{eqnarray}
\Delta_{D}m^2_{\tilde{f}}
\,=\,\frac{1}{5}
     \left(F_{f\,}\alphaone{2}
          +24\,\wt{Y}_{f}\alphafive{2}\right)m_D^2 \ .
\end{eqnarray}
The magnitude of $m_{D}^2$ can be calculated 
when we explicitly fix the model and its breaking pattern.
To be generic, however,
let us take $m_{D}^2$ to be a free parameter.
With the help of $\alphafive{}\fun{M_X}\approx\alphaone{}\fun{M_X}$,
the above expression can be rewritten into
\begin{eqnarray}
\Delta_{D}m^2_{\tilde{f}}
\ \approx\,
   \frac{\sqrt{40}}{5}
   \left(\wt{X}_{f\,}
        +\sqrt{24}\,T^{24}_{\F\,f}\right)\alphaone{2}m_D^2
\,=\,\sqrt{40}\,X_f\VEV{D_\chi} \ ,
\end{eqnarray}
where $X$ is the $U(1)_\chi$ generator (\ref{hypercharge})
in the $SO(10)$ normalization,
and $\VEV{D_\chi}\equiv\alpha^2_5m_D^2$.
Using the value of $U(1)_\chi$ charges 
shown in Table~\ref{table:coefficient2}, we have 
\begin{eqnarray}
\Delta_D m^2_{\tilde L}  ={}-3\VEV{D_\chi} \ , \qquad
\Delta_D m^2_{\tilde e}\,=\,  \VEV{D_\chi} \ ,
\label{Dterm:slepton}
\end{eqnarray}
which give the opposite effect on 
left- and right-handed slepton masses.
Thus it is not possible to 
improve the degeneracy for both of them at the same time
by the $D$-term contribution\rlap.\,\footnote{
$D$-term contribution
through the breaking $E_6 \rightarrow SO(10)$
can be used~\cite{CHK} to increase the slepton masses,
which are originally tachyonic in a scenario of 
anomaly-mediated SUSY breaking~\cite{anom-med}.
Such contributions might also be helpful here
since the broken $U(1)$ charge takes the same sign
for all of quarks and leptons.
}

Eq.~(\ref{LSP}) indicates that
the right-handed sleptons tend to be the LSP.
Therefore a positive contribution
$\Delta_{D}m^2_{\tilde{e}}>0$ is favored to make them heavier,
and this slightly improves 
the degeneracy for the right-handen sleptons.
However,
the degeneracy for left-handed sleptons becomes worse.
{}For instance, we have
$\Delta_{\tilde L} = 2.1 \times 10^{-1}$
if we take $\VEV{D_\chi}= 0.1\Mfive{2}(M_X)$.

%%%%%%%%%%%%%%%%%%%%%%%%%%%%%%%%%%%%%%%%%%%%%%%%%%%%%%%%%%%%%%%%%%%%%%
\subsection{Note on SC threshold effects}
\label{subsec:SCthreshold}
%%%%%%%%%%%%%%%%%%%%%%%%%%%%%%%%%%%%%%%%%%%%%%%%%%%%%%%%%%%%%%%%%%%%%%

When strongly-coupled SC sector decouples around the scale $M_C$,
there arises another type of threshold corrections,
which we refer to as SC threshold effects.
We have little to say about such effects;
one could in principle evaluate SC threshold effects
once the SC sector is specified,
but it requires hard work of understanding strong dynamics.
Basically we do not expect that
Yukawa hierarchy generated by large anomalous dimensions
is modified so much by SC threshold effects,
provided that the SC sector decouples quickly \cite{NS1}.
We have assumed that the same is true
in our estimation of sfermion masses and degeneracy.

There is a special case in which
SC threshold effects are reliably calculated.
Consider for definiteness the MSSM coupled to SC sector,
whose decoupling is caused by invariant mass terms of the order $M_C$.
Suppose also that all the soft SUSY-breaking parameters are given
{\it purely}\/ by anomaly mediation \cite{anom-med}.
Even in this special case,
one can evaluate as before the `convergent value' of sfermion mass,
which is actually the same as anomaly-mediated one calculated 
by using beta-functions of the MSSM coupled to the SC sector.
On the other hand, as was argued in Ref.~\cite{PomarolRattazzi},
the mass parameter $M_C$ should be extended 
to a background (non-dynamical) superfield,
and its $F$-term will affect soft terms. 
One then expects that there arise SC threshold effects such that
after the decoupling, sfermion masses are precisely 
anomaly-mediated one calculated in the MSSM framework.
Thus the SC threshold effects are calculable in this case,
although they are of no interest because of tachyonic sleptons.
Note that the latter sfermion masses and 
the expected convergent values are of the same order of magnitudes,
$m^2_{\tilde{f}}=\order{\alp{a}M_a^2}$.

There is an interesting puzzle here.
Originally we are interested in the SCFT approach
because it can provide degenerate sfermion spectrum 
no matter how SUSY is mediated;
if we consider RG flows of each sfermion mass
corresponding to various initial conditions at the cut-off scale,
all the flows converge on a certain value thanks to the SC dynamics.
Such RG flows would contain a special flow 
corresponding to anomaly-mediated spectrum.
Now, suppose that SUSY breaking is mediated
{\it not}\/ purely by anomaly mediation.
Nevertheless, the above convergence property of RG flows 
would imply that the sfermion spectrum at the decoupling scale,
and thus SC threshold effects, 
be almost the same as anomaly-mediated one.
Specifically the MSSM coupled to SC sector would always 
lead to tachyonic sleptons no matter how SUSY is initially mediated.

The resolution of this apparent puzzle is that
the convergent value of each sfermion mass $m_{\tilde{f}}^2$
is independent of the initial conditions for $m_{\tilde{f}}^2$,
but does depend on the SM gaugino masses.
Therefore sfermion mass spectrum can be different 
from anomaly-mediated one, 
provided that the SM gaugino masses are different.

At any rate, a lesson is that SC threshold effects 
would do affect sfermion masses, 
but their size would be at most comparable 
to the convergent values used in the previous sections.
Therefore we do not expect that
our results would be modified substantially.
It is interesting to confirm this expectation
by explicit calculations.

%%%%%%%%%%%%%%%%%%%%%%%%%%%%%%%%%%%%%%%%%%%%%%%%%%%%%%%%%%%%%%%%%%%%%%
\section{Conclusion and discussion}
\label{sec:conclusion}
\setcounter{footnote}{0}
%%%%%%%%%%%%%%%%%%%%%%%%%%%%%%%%%%%%%%%%%%%%%%%%%%%%%%%%%%%%%%%%%%%%%%

We have studied sfermion mass degeneracy 
within the framework of GUTs coupled to SC sectors.
In the $SU(5)$ model, the degeneracy factor $\Delta_{\tilde e}$ 
for the right-handed sleptons becomes worse than in the MSSM.
The reason is that the right-handed lepton $e_\R^c$ 
is embedded into $10$-dimensional representation and  
the convergent mass value is enlarged.
Models with larger gauge group 
like $SO(10)$ and $E_6$ will have the same feature, 
as long as the SC sector decouples 
at the scale larger than the GUT breaking scale.
One way to keep $e_\R^c$ a non-Abelian singlet
is the flipped $SU(5)_\F\times U(1)_\F$ model.
In this model, the degeneracy of the right-handed sleptons
can be improved if the $U(1)_\F$ gaugino mass $\Mone{}$ 
is smaller than the $SU(5)_\F$ one;
a small $\Mone{}$ reduces the convergent value
and the mass difference,
while the $SU(5)_\F$ component of the bino mass
enhances the averaged slepton mass.

In this paper,
we have estimated sfermion mass degeneracy
without specifying the model for the SC sector.
To this end, we have assumed that 
the convergent value of sfermion masses at the decoupling scale 
is dominately determined by the SM one-loop terms in the RGEs.
This assumption is plausible for the $SU(5)$ case,
but is crucial especially for the right-handed sleptons
in the MSSM and the flipped model.
A clever model building will be required 
when the SC sector is vector-like under the SM-sector gauge group.
We have also assumed that our estimation is not 
substantially modified by SC threshold effects.
We expect that the size of such effects will be at most 
comparable to the estimated convergent value of sfermion masses.
It is worth while examining these points by explicit calculations.

{}Finally let us speculate about 
possible extensions of the present work.
We have neglected effects from the third family 
of quarks and leptons and their Yukawa couplings.
In quark sector, it will be in safe to neglect the third family,
since they have only small mixings with the others
in many realistic Yukawa matrices.
However, the same is not true in lepton sector.
In particular, the flipped $SU(5) \times U(1)$ model requires that 
the tau neutrino Yukawa coupling is 
as large as the top Yukawa coupling.
In general, the large mixings in lepton sector 
induce a significant flavor violation $\delta^\ell_{12}$ 
through the radiative corrections above the mass scale $M_\nu$ 
of the right-handed neutrinos~\cite{BM,HMTY,ST,KKST},
and that constrains sfermion mass spectrum and/or 
requires a specific form of Yukawa matrices.
{}From this viewpoint, an interesting possibility is that
a proper coupling to SC sector and subsequent IR convergence property
can eliminate such flavor violation as well
if $M_C$ is taken below $M_\nu=10^{13}\,\hbox{--}\,10^{15}$~GeV.
We need further study 
concerning neutrino Yukawa couplings\rlap.\,\footnote{
An application of the SC idea 
to the neutrino as well as Higgs sectors
was discussed in Ref.~\cite{KuboSuematsu}.
}

In models with a product gauge group like $SU(5)\times U(1)$,
the degeneracy can be improved 
if the gaugino masses do not satisfy the usual GUT relation
so that the convergent value of sfermion mass is much suppressed
compared with the radiative corrections in the MSSM regime.
The $SU(5)\times{}U(3)_{{\rm H}}$ models~\cite{hyperGUT} 
will have a similar property.
We speculate that
even with a simple gauge group $SU(5)$,
such a suppression of the convergent value could be realized
if one supposes that the $SU(5)$ gauge multiplets
live in an extra-dimensional space-times.
In this setup, the power-law evolution of 
gauge coupling and gaugino mass~\cite{power-law,kkmz,ktz}
makes them drastically small above $M_X$
(in such a way that $M_5/\alpha_5$ is still RG invariant~\cite{kkmz})
provided that the gauge coupling is asymptotically free.
Consequently the degeneracy would be improved 
if $M_C$ is taken as the energy scale where
$\alpha_5$ and $M_5$ are very small.
To keep our RG calculations reliable in the SC regime,
quarks and leptons as well as SC-sector fields 
are supposed to live in four dimensions.
Such a setup could also explain why 
our gauge coupling $\alpha_5$ is smaller than that in the SC sector, 
thanks to extra-dimensional volume suppression.

The Nelson-Strassler scenario, which we have focused in this paper,
assumes that anomalous dimensions of quarks and leptons 
take flavor-dependent values at the SC fixed point.
The difference between $\gamma_i^*$ and $\gamma_j^*$ 
leads to hierarchical Yukawa matrices, but at the same time, 
the difference between $\Gamma_i$ and $\Gamma_j$
leads to a distinctive non-degeneracy of sfermion masses.
In the `Yukawa hierarchy transfer' scenario~\cite{KNT}, 
the SC dynamics is supposed to be flavor-blind at the fixed point,
$\gamma_i^* = \gamma_j^*$ and $\Gamma_i = \Gamma_j$,
ensuring complete degeneracy of sfermion masses.
The hierarchy among Yukawa couplings $y_{ij}$ is derived 
by assuming the inverse hierarchy in the {\it messenger}\/ couplings 
$\lambda_{irs}$ at the cut-off scale and 
by {\it transferring}\/ the initial hierarchy through the SC dynamics.
Moreover, the assumed hierarchy can be derived, \eg, 
through the FN mechanism in the SC sector
{\it without}\/ spoiling the sfermion degeneracy
by family-dependent $D$-term contributions.
This new scenario provides an alternative way of 
realizing degenerate sfermion spectrum
and hierarchical Yukawa matrices at the same time.

\section*{Acknowledgments}

The authors wold like to thank J.~Kubo, N.~Maekawa, J.~Sato, 
D.~Suematsu, M.~Tanimoto and T.~Yanagida for useful discussions.
We would like to express our thanks to
the organizers of Summer Institute~2001, Yamanashi,
where a part of the work was done.

%%%%%%%%%%%%%%%%%%%%%%%%%%%%%%%%%%%%%%%%%%%%%%%%%%%%%%%%%%%%%%%%%%%%%%

\end{document}